\documentstyle[osa,epsf,rotate,manuscript]{revtex} 
\newcommand{\dfrac}{\displaystyle\frac}
\begin{document}
\title{\large {\bf The surface diffuseness and the spin-orbital splitting
in relativistic continuum Hartree-Bogoliubov theory }}
\author{J. Meng and I.Tanihata$^{a}$}
\address{$^{*}$Department of Technical Physics, Peking University,\\
Beijing 100871, P.R. China \\
$^{a}$The Institute of Physical and Chemical Research (RIKEN)  \\
Hirosawa 2-1, Wako-shi, Saitama, 351-0198 JAPAN \\ 
$^{b}$Institute of Modern Physics, Chinese Academy of Sciences \\
Lanzhou 730000,  China  \\
         e-mail: meng@rikaxp.riken.go.jp}
\date{\today}
\maketitle
\begin{abstract}
The Relativistic Continuum
Hartree Bogoliubov theory ( RCHB ), which is
the extension of the
Relativistic Mean Field and the Bogoliubov transformation
in the coordinate representation, has been
used to study tin isotopes. 
The pairing correlation is
taken into account by a density-dependent force of zero
range. RCHB is used to describe the even-even tin
isotopes all the way from the proton drip line
to the neutron drip line. The contribution of the
continuum which is important for nuclei near the drip-line 
has been taken into account. The theoretical $S_{2n}$ as well as
the neutron, proton, and matter $rms$ radii are presented
and compared with the experimental values where they 
exist. The change of the potential surface with the
neutron number has been investigated.
The diffuseness of the potentials in tin isotopes is analyzed through the
spin-orbital splitting in order to provide new way
to understand the halo phenomena in exotic nuclei. 
The systematic of the isospin  and energy dependence of
these results are extracted and analyzed.
\end{abstract}
\par


\baselineskip = 24pt


\section{Introduction}     

Highly unstable nuclei with extreme proton and neutron
ratio are now accessible with the help of the radioactive nuclear beam
facilities.
The extreme proton and neutron ratio of these nuclei and physics
connected with these unprecedented low density matter have
attracted worldwide attention
in nuclear physics as well as other fields such as astrophysics.
The recent research in this topic includes the detailed 
structure, its mode of excitation, and the reaction mechanism. 
Using the measured interaction cross sections it is possible to
extract the nuclear root mean-square ($rms$) matter radii, the
nuclear mass density and the other related observations.
In  these investigations, the observed
sudden rise in the measured interaction cross-sections in the
neutron rich light nuclei has been
attributed to the corresponding large increase in the nuclear
$rms$ matter radii. This fact is an indication of the sudden change
in the structure of these nuclei due to the addition of the last few 
neutrons. The celebrated example
in these investigations is the discovery of the
phenomenon of "neutron halo" \cite{THH.85b}, for example in $^{11}$Li.
The neutron halo is a state in which neutrons spread like a thin mist 
around the nucleus. It is qualitatively associated with the very small
separation energy ($S_n$) of the last one or two neutrons. 
Due to this small value
of $S_n$ in such nuclei, the tail of its last neutron
wavefunction and so also the corresponding density
spreads far out from the center of the nucleus. 

New theoretical models and techniques are being developed 
in order to describe unique phenomena in exotic nuclei far
from the line of stability.  Particularly, on the 
neutron-rich side, the property of exotic nuclei 
involves the weak binding of the outermost 
neutrons, the coupling of bound state and the continuum, 
the diffuse particle distribution and isospin 
dependence of the shell structure.  Due to their 
relevance to the r-process in nucleosynthesis, nuclei 
near the drip-line are also very important in nuclear astrophysics.
As for the theoretical microscopic description of unstable nuclei is
concerned, so far rather different techniques have been used, 
e.g., the Relativistic Continuum Hartree-Bogoliubov ( RCHB ) 
theory \cite{MR.96,MR.97,ME.98};
the exact solution of few-body equations treating inert sub-clusters
as point particles\cite{Zhu.93};
the Skyrme Hartree-Fock-Bogoliubov (HFB) theory
\cite{DFT.84,Ter.96};  and two-frequency
shell model \cite{KUO.97}.

The few body method and
a shell model diagonalization start by fitting
the ground state properties of the nuclei or neighboring nuclei
and aim at predicting the excited state properties. 
Whereas the relativistic and
conventional mean field approaches are based on the description of global
experimental data for nuclei throughout the nuclear chart.
The principal goal here is to obtain a fair description of bulk
properties, particularly for the ground states. The mean field
has certain advantages in describing and predicting
quantities, such as the radii, mass, tunneling probabilities,
etc., in stable nuclei and their future extendability to exotic nuclei
when the continuum is properly taken into account.

The RCHB theory, which is the extension of the
Relativistic Mean Field ( RMF ) and the Bogoliubov transformation
in the coordinate representation, provides not only a unified 
description of mean field and pairing correlation 
but also the proper description for the continuum and the 
coupling between the bound state and the continuum \cite{ME.98}.

Recently, a fully self-consistent description of the chain 
of Lithium isotopes ranging
from $^{6}$Li to $^{11}$Li has been given by the 
RCHB \cite{MR.96}. The  halo in $^{11}$Li has 
been successfully reproduced in this
self-consistent picture. Excellent agreement with recent
experimental data on the radii, binding energy and 
density distribution is obtained. This
remarkable success is mainly attributed to the proper treatment
of the continuum states and the pairing by solving the RCHB equations.
This procedure helps to incorporate
correctly the scattering of
Cooper pairs to the $2s_{1/2}$ level, which is more close to the
threshold than the $1d_{5/2}$ and $1d_{3/2}$ levels, in the continuum. 
The contribution from the continuum is proved to be crucial. 
Therefore the relativistic mean field provides another picture
of halo in nuclei: Due to the extreme low proton to neutron ratio,
the surface of the neutron potentials become highly diffuse.
Because the orbital with small orbital angular momentum
has less centrifugal barrier, the particles with weak binding
or even in the continuum in these orbital
will have more chance to tunnel.
In order to study the influence of correlation and
many-body effects, it would be very interesting to find also
nuclei with a larger number of neutron distributed in the
halo. Based on the RCHB, a new phenomenon "Giant Halo" 
has been predicted 
in the Zr nuclei close to the neutron drip line, which
are composed not only by one or two neutrons, as is the
case in the halos investigated so far in light $p$-shell
nuclei, but which contain up to 6 neutrons \cite{MR.97}. 

The development of a proton skin as well as neutron skin
in Na isotopes has been systematically studied
with RCHB in Ref. \cite{MTY.97}, where the pairing and blocking
effect have been treated self-consistently.
A Glauber model calculation
has been carried out with the density obtained from RCHB.
A good agreement has been obtained with the measured
cross sections
for $^{12}$C as a target and a rapid
increase of the cross sections has been predicted for
neutron rich Na isotopes beyond $^{32}$Na.
After systematic examination of
the neutron, proton and matter distributions
in the Na nuclei from the proton drip-line
to the neutron drip-line,
a relation between the tail part of the density and the
shell structure has been found.
The tail of the matter distribution is not so sensitive
to how many particles are filled in a major shell. Instead
it is very sensitive to whether this shell has been occupied or
not.  The physics behind the skin and halo has been revealed as
a spatial demonstration of shell effect: simply the extra
neutrons are filled in the next shell and sub-shell. This is in
agreement with the mechanism observed so far in the halo system but
more general. 

From the mean field point of view, the properties of 
all the nucleons in the nuclei are determined by the 
mean potential provided by their interaction with the 
other nucleons. Therefore the study of the isospin dependence
of the potential, which become highly diffuse near the
particle drip line, is crucial to understand
unstable nuclei and the study of the surface diffuseness
will provide another mean to understand exotic nuclei. 
In the present work, we extend our former investigation and 
study the surface diffuseness, the isospin dependence
of the potential,  and spin-orbital splitting in tin isotopes. 
For exotic nuclei, as orbitals with small orbital angular momentum
near the threshold will extend far out from the nuclear center, 
and thus the pairing correlation can couple the
bound states and  positive energy states.
Particularly on the neutron-rich
side, as neutron carries no electric charge, the
neutron drip line is located very far from the valley of
$\beta$-stability. It is crucial to understand how the isospin dependence
of the potential and the diffuseness develops 
in unstable nuclei.
The surface diffuseness can be strongly related 
with the spin-orbital splitting, which could be measured 
experimentally.  

In the present
study, isospin dependence of the potential diffuseness, 
their relation with the spin-orbital splitting , and the 
ground-state properties of drip-line systems are investigated
by means of the self-consistent RCHB approach.
The shell structure and its isospin dependence in
tin isotope is discussed in the canonical basis.
The isospin-dependence and energy-dependence of the
neutron spin-orbit splitting are discussed.
An outline of the RCHB formalism is briefly reviewed 
in Sec. II.
In Sec. III, the isospin-dependence and energy-dependence of the 
neutron spin-orbit splitting,
the neutron, proton, and matter density distributions
are presented for Sn isotopes.
A brief summary is given in the last section.


\section{RCHB Theory}

The RCHB theory, which is the extension of the
RMF and the Bogoliubov transformation
in the coordinate representation, is suggested in Ref. \cite{MR.96}, its
detailed formalism and numerical
solution can be found in Ref. \cite{ME.98} and the references therein.
The basic ansatz of the RMF theory is a Lagrangian
density whereby nucleons are described
as Dirac particles which interact via the exchange of various
mesons and the photon.
The mesons considered are the scalar
sigma ($\sigma$), vector omega ($\bf \omega$) and iso-vector vector rho
($\bf \vec \rho$).  The rho
($\bf \vec \rho$) meson provides the necessary isospin
asymmetry.
The scalar sigma meson moves in a self-interacting field having cubic and
quadratic terms with strengths $g_2$ and $g_3$ respectively.
The Lagrangian then consists of the free baryon
and meson parts and the interaction part with minimal coupling,
together with the nucleon mass $M$, and $m_\sigma$,
$g_\sigma$, $m_\omega$, $g_\omega$, $m_\rho$, $g_\rho$  the masses and
coupling constants of the respective mesons and:
\begin{eqnarray}
\begin{array}{cc}
{\cal L} &= \bar \psi (i\rlap{/}\partial -M) \psi +
        \,{1\over2}\partial_\mu\sigma\partial^\mu\sigma-U(\sigma)
        -{1\over4}\Omega_{\mu\nu}\Omega^{\mu\nu} \\
        \ &+ {1\over2}m_\omega^2\omega_\mu\omega^\mu
        -{1\over4}{\vec R}_{\mu\nu}{\vec R}^{\mu\nu} +
        {1\over2}m_{\rho}^{2} \vec\rho_\mu\vec\rho^\mu
        -{1\over4}F_{\mu\nu}F^{\mu\nu} \\
        & -  g_{\sigma}\bar\psi \sigma \psi~
        -~g_{\omega}\bar\psi \rlap{/}\omega \psi~
        -~g_{\rho}  \bar\psi
        \rlap{/}\vec\rho
        \vec\tau \psi
        -~e \bar\psi \rlap{/}A \psi.
\label{Lagrangian}
\end{array}
\end{eqnarray}

The field tensors for the
vector mesons are given as:
\begin{eqnarray}
\left\{
\begin{array}{lll}
   \Omega^{\mu\nu}   &=& \partial^\mu\omega^\nu-\partial^\nu\omega^\mu, \\
   {\vec R}^{\mu\nu} &=& \partial^\mu{\vec \rho}^\nu
                        -\partial^\nu{\vec \rho}^\mu
                        - g^{\rho} ( {\vec \rho}^\mu
                           \times {\vec \rho}^\nu ), \\
   F^{\mu\nu}        &=& \partial^\mu \vec A^\nu-\partial^\nu \vec A^\mu.
\end{array}   \right.
\label{tensors}
\end{eqnarray}
For a realistic description of
nuclear properties a nonlinear self-coupling for the scalar
mesons turns crucial \cite{BB.77}:
\begin{equation}
   U(\sigma)~=~\dfrac{1}{2} m^2_\sigma \sigma^2_{}
            ~+~\dfrac{g_2}{3}\sigma^3_{}~+~\dfrac{g_3}{4}\sigma^4_{}
\end{equation}

The classical variation principle gives the following equations of
motion :
\begin{equation}
   [ {\vec \alpha} \cdot {\vec p} +
     V_V ( {\vec r} ) + \beta ( M + V_S ( {\vec r} ) ) ]
     \psi_i ~=~ \epsilon_i\psi_i
\label{spinor1}
\end{equation}
for the nucleon spinors and
\begin{eqnarray}
\left\{
\begin{array}{lll}
  \left( -\Delta \sigma ~+~U'(\sigma) \right ) &=& g_\sigma\rho_s     
\\
   \left( -\Delta~+~m_\omega^2\right )\omega^{\mu} &=&
                    g_\omega j^{\mu} ( {\vec r} )
\\
   \left( -\Delta~+~m_\rho^2\right) {\vec \rho}^{\mu}&=&
                    g_\rho \vec j^{\mu}( {\vec r} )
\\
          -\Delta~ A_0^{\mu} ( {\vec r} ) ~ &=&
                           e j_{\rho}^{\mu}( {\vec r} )
\end{array}  \right.
\label{mesonmotion}
\end{eqnarray}
for the mesons, where
\begin{eqnarray}
\left\{
\begin{array}{lll}
   V_V( {\vec r} ) &=&
      g_\omega\rlap{/}\omega + g_\rho\rlap{/}\vec\rho\vec\tau
         + \dfrac{1}{2}e(1-\tau_3)\rlap{\,/}\vec A , \\
   V_S( {\vec r} ) &=&
      g_\sigma \sigma( {\vec r} ) \\
\end{array}
\right.
\label{vaspot}
\end{eqnarray}
are the vector and scalar potentials respectively and
the source terms for the mesons are
\begin{eqnarray}
\left\{
\begin{array}{lll}
   \rho_s &=& \sum_{i=1}^A \bar\psi_i \psi_i
\\
   j^{\mu} ( {\vec r} ) &=&
               \sum_{i=1}^A \bar \psi_i \gamma^{\mu} \psi_i
\\
   \vec j^{\mu}( {\vec r} ) &=&
          \sum_{i=1}^A \bar \psi_i \gamma^{\mu} \vec \tau  \psi_i
\\
   j^{\mu}_p ( {\vec r} ) &=&
      \sum_{i=1}^A \bar \psi_i \gamma^{\mu} \dfrac {1 - \tau_3} 2  \psi_i,
\end{array}  \right.
\label{mesonsource}
\end{eqnarray}
where the summations are over the valence nucleons only.
It should be noted that as usual, the present approach
neglects the contribution of
negative energy states, i.e.,  no-sea approximation,
which means that the vacuum
is not polarized. The coupled equations Eq.(\ref{spinor1})
and Eq.(\ref{mesonmotion}) are nonlinear quantum
field equations, and their exact solutions are very complicated.
Thus the mean field approximation is generally used: i.e.,
the meson field operators in Eq.(\ref{spinor1}) are replaced
by their expectation values. Then the nucleons move
independently in the classical meson fields. The
coupled equations are self-consistently solved by iteration.
\par

For spherical nuclei, i.e., the systems with rotational symmetry,
the potential of the nucleon and the sources of
meson fields depend only on the radial coordinate $r$. The spinor
is characterized by the angular momentum quantum numbers $l$, $j$,$m$, the
isospin $t = \pm \dfrac 1 2$ for neutron and proton respectively, and the
other quantum number $i$. The Dirac spinor has the form:
\begin{equation}
   \psi ( \vec r ) =
      \left( { {\mbox{i}  \dfrac {G_i^{lj}(r)} r {Y^l _{jm} (\theta,\phi)} }
      \atop
       { \dfrac {F_i^{lj}(r)} r (\vec\sigma \cdot \hat {\vec r} )
       {Y^l _{jm} (\theta,\phi)} } }
      \right) \chi _{t}(t),
\label{reppsi}
\end{equation}
where $Y^l _{jm} (\theta,\phi)$ are the spinor spherical harmonics
and $G_i^{lj}(r)$ and $F_i^{lj}(r)$ are the remaining radial
wave function for upper and lower components. They are normalized
according to
\begin{equation}
\int_0^{\infty}dr ( | G_i^{lj}(r) |^2 + | F_i^{lj}(r) |^2 ) = 1.
\end{equation}
The radial equation of spinor Eq. (\ref{spinor1}) can be reduced as :
\begin{eqnarray}
\left\{
\begin{array}{lll}
   \epsilon_i G_i^{lj}(r) &=& ( - \dfrac {\partial} {\partial r}
      + \dfrac {\kappa_i} r )  F_i^{lj}(r) + ( M + V_S(r) + V_V(r) ) G_i^{lj}(r)
\\
   \epsilon_i F_i^{lj}(r) &=& ( + \dfrac {\partial} {\partial r}
      + \dfrac {\kappa_i} r )  G_i^{lj}(r)
      - ( M + V_S(r) - V_V(r) ) F_i^{lj}(r) .
\end{array}  \right.
\label{spinorradical}
\end{eqnarray}
where 
\begin{displaymath}
   \kappa =
      \left\{
         \begin{array}{ll}
            -(j+1/2)  & for ~ j=l+1/2 \\
            +(j+1/2)  & for ~ j=l-1/2. \\
         \end{array}
      \right.
\end{displaymath}
The meson field equations become simply radical Laplace equations
of the form:
\begin{equation}
\left( \frac {\partial^2} {\partial r^2}  - \frac 2 r
\frac  {\partial}   {\partial r} + m_{\phi}^2 \right)\phi = s_{\phi} (r),
\label{Ramesonmotion}
\end{equation}
$m_{\phi}$ are the meson masses for $\phi = \sigma, \omega,\rho$
and for photon ( $m_{\phi} = 0$ ). The source terms are:
\begin{eqnarray}
s_{\phi} (r) = \left\{
\begin{array}{ll}
-g_\sigma\rho_s - g_2 \sigma^2(r)  - g_3 \sigma^3(r)
& { \rm for ~ the ~  \sigma~  field } \\
g_\omega \rho_v    & {\rm for ~ the ~ \omega ~ field} \\
g_{\rho}  \rho_3(r)       & {\rm for~ the~ \rho~ field} \\
e \rho_c(r)  & {\rm for~ the~ Coulomb~ field}, \\
\end{array}
\right.
\end{eqnarray}

\begin{eqnarray}
\left\{
\begin{array}{lll}
   4\pi r^2 \rho_s (r) &=& \sum_{i=1}^A ( |G_i(r)|^2 - |F_i(r)|^2 ) \\
   4\pi r^2 \rho_v (r) &=& \sum_{i=1}^A ( |G_i(r)|^2 + |F_i(r)|^2 ) \\
   4\pi r^2 \rho_3 (r) &=& \sum_{p=1}^Z ( |G_p(r)|^2 + |F_p(r)|^2 )
               -  \sum_{n=1}^N ( |G_n(r)|^2 + |F_n(r)|^2 ) \\
   4\pi r^2 \rho_c (r) &=& \sum_{p=1}^Z ( |G_p(r)|^2 + |F_p(r)|^2 ) . \\
\end{array}
\right.
\label{mesonsourceS}
\end{eqnarray}
The Laplace equation can in principle be solved by the Green function:
\begin{equation}
\phi (r) = \int_0^{\infty} r'^2 dr' G_{\phi} (r,r') s_{\phi}(r'),
\end{equation}
where for massive fields
\begin{equation}
G_{\phi} (r,r') = \frac 1 {2m_{\phi}} \frac 1 {rr'}
( e^{-m_{\phi} | r-r'|} -  e^{-m_{\phi} | r+r'|} )
\end{equation}
and for Coulomb field
\begin{equation}
G_{\phi} (r,r') =
\left\{
\begin{array}{ll}
1/r  & {\rm for ~ r~ >~ r' } \\
1/r' & {\rm for ~r~ <~ r'} .\\
\end{array}
\right.
\end{equation}

The Eqs.(\ref{spinorradical}) and (\ref{Ramesonmotion})could be solved 
self-consistently in the usual RMF approximation.  
For RMF, however, as the classical meson fields are used, 
the equations of motion
for nucleons derived from Eq.(\ref{Lagrangian}) do not contain
pairing interaction. In order to have pairing interaction, one has to 
quantize the meson fields which leads to a Hamiltonian with two-body 
interaction. Following the standard procedure of 
Bogoliubov transformation, a Dirac Hartree-Bogoliubov equation 
could be derived and then a unified description of the mean 
field and pairing correlation in nuclei 
could be achieved. For the details, see 
Ref. \cite{ME.98} and the references therein. 
The RHB equations are as following:
\begin{equation}
   \int d^3r'
   \left( \begin{array}{cc}
          h-\lambda &   \Delta \\
          \Delta    &  - h+\lambda
          \end{array} \right) 
   \left( { \psi_U \atop\psi_V } \right)  ~
   = ~ E ~ \left( { \psi_U \atop \psi_V } \right), 
\label{ghfb}
\end{equation}
where 
\begin{equation}
   h(\vec r,\vec r') =  \left[ {\vec \alpha} \cdot {\vec p} +
      V_V ( {\vec r} ) + \beta ( M + V_S ( {\vec r} ) ) \right]
      \delta(\vec r,\vec r')
\label{NHamiltonian}
\end{equation}
is the Dirac Hamiltonian and the Fock term has been neglected as is 
usually done in RMF. The pairing potential is :
\begin{eqnarray}
    \Delta_{kk'}(\vec r, \vec r')
    &=& \int d^3r_1 \int d^3r_1' \sum_{\tilde k \tilde k'}
    V_{kk',\tilde k \tilde k'} ( \vec r \vec r'; \vec r_1 \vec r_1' )
    \kappa_{\tilde k \tilde k'}  (\vec r_1, \vec r_1' ).
\label{gap}
\end{eqnarray}
It is obtained from one-meson exchange interaction
$V_{kk',\tilde k \tilde k'} (\vec r \vec r'; \vec r_1 \vec r_1' )$ in the
$pp$-channel and the pairing tensor $\kappa=V^*U^T$ 
\begin{eqnarray}
   \kappa_{k k'}( \vec r, \vec r') = < | a_{k} a _{k'} | >
   = \psi_V^{k}(\vec r) ^* \psi_U^{k'}(\vec r)^T
\end{eqnarray}
The nuclear density is as following:
\begin{eqnarray}
   \rho(\vec r ,\vec r' ) 
   = \sum_i^{lj}  \psi_V^{ilj} (\vec r) ^* \psi_V ^{ilj} (\vec r ' ) .
\end{eqnarray}
As in Ref. \cite{ME.98}, $V_{kk',\tilde k \tilde k'}$ used for the 
pairing potential in Eq.(\ref{gap}) is either the density-dependent 
two-body force of zero range 
with the interaction strength $V_0$ and the nuclear matter density $\rho_0$:
\begin{equation}
   V(\mbox{\boldmath $r$}_1,\mbox{\boldmath $r$}_2) = V_0
     \delta(\mbox{\boldmath $r$}_1-\mbox{\boldmath $r$}_2)
     \frac{1}{4}\left[1-
     \mbox{\boldmath $\sigma$}_1\mbox{\boldmath $\sigma$}_2\right]
     \left(1 - \frac{\rho(r)}{\rho_0}\right)
\label{vpp}
\end{equation}
or Gogny-type finite range force 
with the parameter $\mu_i$, $W_i$, $B_i$, $H_i$ and
$M_i$ ($i=1,2$) as the finite range part of the Gogny
force \cite{BGG.84}:
\begin{equation}
   V(\mbox{\boldmath $r$}_1,\mbox{\boldmath $r$}_2)
      ~=~\sum_{i=1,2}
       e^{((\mbox{\boldmath $r$}_1-\mbox{\boldmath$r$}_2) / \mu_i)^2}
       (W_i + B_i P^{\sigma} - H_i P^{\tau} - M_i P^{\sigma} P^{\tau})
\label{vpp2}
\end{equation}
A Lagrange multiplier $\lambda$ is introduced to fix the 
particle number for the neutron
and proton as $N = \mbox {Tr} \rho _n$ and  $Z = \mbox {Tr} \rho _p$ .

In order to describe both the continuum and the bound states 
self-consistently, the RHB theory must be sovled in  
coordinate representation, i.e., the.
Relativistic Continuum Hartree-Bogoliubov ( RCHB )
theory \cite{ME.98}. It is then applicable to both 
exotic nuclei and normal nuclei.
In Eq. (\ref{ghfb}), the spectrum of the system is unbound from 
above and from below the Fermi surface, and the
eigenstates occur in pairs of opposite energies. 
When spherical symmetry is imposed on the solution of the
RCHB equations, the wave function can be conveniently
written as
\begin{equation}
   \psi^i_U = 
      \left( {\displaystyle {\mbox{i} \frac {G_U^{ilj}(r)} r }  \atop
     {\displaystyle \frac {F_U^{ilj}(r)} r 
        (\vec\sigma \cdot \hat {\vec r} )  } }
              \right) {Y^l _{jm} (\theta,\phi)}  \chi_{t}(t) ,  
   \psi^i_V =
       \left( {\displaystyle {\mbox{i} \frac {G_V^{ilj}(r)} r }  \atop
          {\displaystyle \frac {F_V^{ilj}(r)} r 
             (\vec\sigma \cdot \hat {\vec r} )
       } } \right)  {Y^l _{jm} (\theta,\phi)}  \chi_{t}(t).
\end{equation}

The above equation Eq.(\ref{ghfb})  depends only 
on the radial coordinates and can be
expressed as the following integro-differential equation:
\begin{eqnarray}
\left\{
   \begin{array}{lll}
      \displaystyle
      \frac {d G_U(r)} {dr} + \frac {\kappa} r G_U(r) -
       ( E + \lambda-V_V(r) + V_S(r) ) F_U(r) +
         r \int r'dr' \Delta(r,r')  F_V(r') &=& 0  \\
      \displaystyle
      \frac {d F_U(r)} {dr} - \frac {\kappa} r F_U(r) +
       ( E + \lambda-V_V(r)-V_S(r) ) G_U(r) +
         r \int r'dr' \Delta(r,r') G_V(r') &=& 0 \\
      \displaystyle
      \frac {d G_V(r)} {dr} + \frac {\kappa} r G_V(r) +
       ( E - \lambda+V_V(r)-V_S(r) ) F_V(r) +
         r \int r'dr' \Delta(r,r') F_U(r') &=& 0 \\
      \displaystyle
      \frac {d F_V(r)} {dr} - \frac {\kappa} r F_V(r) -
       ( E - \lambda+V_V(r)+V_S(r) ) G_V(r) +
         r \int r'dr' \Delta(r,r')  G_U(r') &=& 0, \\
\end{array}
\right.
\label{CoupEq}
\end{eqnarray}
where the nucleon mass is included in the 
scalar potential $V_S(r)$.
Eq.(\ref{CoupEq}), in the case of $\delta$-force Eq.(\ref{vpp}),
is reduced to normal coupled differential equations and can be solved
with shooting method by Runge-Kutta algorithms.
For the case of Gogny force, the coupled integro-differential equations 
are discretized in the space and solved by the 
finite element methods, see Ref. \cite{ME.98}
Instead of solving Eqs.(\ref{spinorradical}) and (\ref{Ramesonmotion})
self-consistently for the RMF case, now we have to solve 
Eqs.({\ref{CoupEq}) and (\ref{Ramesonmotion})
self-consistently for the RCHB case. 
As the calculation for Gogny force is very time-consuming,  
we use them for one nucleus and fixed the interaction 
strength in $\delta$-force Eq.(\ref{vpp}).


\section{Results and discussion}

The procedure to solve the RCHB equations 
is the same as in Ref. \cite{ME.98}, we
solve the RCHB equations in a box of the size $R = 25$ fm and
a step size of $0.1$ fm for the
parameter set NLSH \cite{SNR.93} self-consistently.
The $\delta$-force is used in the pairing channel and its 
strength is properly fixed by the Gogny force as in Ref. \cite{ME.98}.
The number of continuum taken into account is decided by 
a cutoff energy, i.e., only the levels lying within 120 MeV
from the Fermi level are taken into account.
For $\rho_0$ we use the nuclear
matter density 0.152 fm$^{-3}$.
The strength $V_0$ of Eq.(\ref{vpp})
is determined by adjusting the corresponding pairing energy
$-\frac{1}{2}\mbox{Tr}\Delta\kappa$ to that of
the Gogny force D1S \cite{BGG.84} in $^{150}$Sn.

In Fig. 1, we show the two neutron separation energies $S_{2n}$ 
of even tin isotopes as a function of the neutron number $N$ 
from the proton drip-line to the neutron drip-line,
including the experimental data ( solid points ), and RCHB with
$\delta$-force ( open circles ). 
The RCHB calculation reproduces the experimental data quite well.
They show a strong kink at $N=50$ and $N=82$. 
The drip line nucleus is predicted at
$^{176}$Sn in the present calculation. 
The available empirical data of $S_{2n}$ have been well 
reproduced. 

The magnitude of the kink at $N=82$ shows that 
the shell effects in the RMF theory are observed to be strong at $N=82$.
Measurement of the mass of $^{134}$Sn, in future, would clarify
the nature of the shell effects at $N=82$. Apart from the kinks 
observed at $N=50$ and $N=82$, $S_{2n}$ gradually decrease 
with $N$. The $S_{2n}$ near drip line nuclei 
$^{176}$Sn decreases relatively faster than 
the case in light nuclei, e.g., Na isotopes \cite{MTY.97}.
This is due to the 
effects of centrifugal barrier. In Sn isotopes, the orbital 
$1i_{13/2}$ is just above the threshold for nuclei near the 
drip line. Due to its big centrifugal barrier, particles filled 
in $1i_{13/2}$ do not contribute to the binding, although the 
potential become highly diffuse. In light nuclei like 
Na isotopes, the orbital
$2p_{1/2}$  and $2p_{3/2}$ are just above the threshold. 
They gain binding energy due to the diffuseness of potential 
with $N$. Therefore the total binding energies of the Na isotopes 
remain unchanged with $N$ near the drip line. 
 
The neutron, proton, matter, and charge radii 
for all even Sn isotopes are given 
in Table 1. The neutron and matter radii increase with the neutron  
number. Also the proton radii increases slowly 
after $^{100}$Sn. 

Encouraged by the success in describing the $S_{2n}$ for tin 
isotopes, we proceed to examine the spin-orbital splitting for the 
whole isotopes. As an example, spin-orbital splitting
\begin{equation}
   E_{ls} = \displaystyle \frac {E_{lj=l-1/2}-E_{lj=l+1/2}} {2l+1}
\label{E-ls}
\end{equation}
versus the binding energy:
\begin{equation}
   E = \displaystyle \frac { ( l+1 ) E_{lj=l-1/2}+ l E_{lj=l+1/2}} {2l+1}
\label{abe}
\end{equation}
in $^{110}$Sn, $^{120}$Sn, $^{130}$Sn, $^{140}$Sn, $^{150}$Sn, 
$^{160}$Sn, and $^{170}$Sn are given in Fig. 2 for the neutron 
spin-orbital partners
($1d_{3/2}, 1d_{5/2}$),($1g_{7/2}, 1g_{9/2}$), 
($1i_{11/2}, 1i_{13/2}$),
($1p_{1/2}, 1p_{3/2}$), ($1f_{5/2}, 1f_{7/2}$), 
and ($1h_{9/2}, 1h_{11/2}$), and the proton 
spin-orbital partners
($1d_{3/2}, 1d_{5/2}$), and  ($1f_{5/2}, 1f_{7/2}$). 
It is very interesting to see that the spin-orbital splitting 
for the neutron and proton is very close to each other, at least 
for ($1d_{3/2}, 1d_{5/2}$), and  ($1f_{5/2}, 1f_{7/2}$) cases.
The splitting decreases monotonically from the proton drip line to the
neutron drip line for all the partners. To see the underlying reason  
of this behavior, it is very helpful to examine the origin of the
spin-orbital splitting in the Dirac equation. For the Dirac nucleon 
moving in a scalar and vector potentials, its equation of motion 
could be de-coupled and reduced for the upper component and 
the lower component, respectively. If it is reduced in 
the lower component, it will be related with another interesting 
topic -- the pseudo-spin symmetry discussed in 
Ref. \cite{MSY.98} and the references therein.
For the moment, we are interested in the spin-orbital splitting, 
the Dirac equation can be reduced for the upper 
component as following: 
\begin{eqnarray}
   & &  [ \frac {d^2} {dr^2}  - \frac 1  {E + 2M - V_V  + V_S}
        \frac {d(2M-V_V+V_S)} {dr} \frac {d} {dr} ] G^{lj}_i(r) \nonumber \\
   &-&  [  \frac { \kappa ( 1 + \kappa ) }  {r^2}
        - \frac 1 {E + 2M - V_V + V_S} \frac {\kappa} r
        \frac {d(2M - V_V + V_S)} {dr}  ] G^{lj}_i (r) \nonumber \\
 = &-&  (E + 2M - V_V + V_S ) ( E - V_V - V_S ) G^{lj}_i (r)
\label{larspinor4}
\end{eqnarray}
where
\begin{eqnarray}
       \kappa =
        \left\{ \begin{array}{cc}
            -l-1,  &  j=l+1/2  \\
            l,     &  j=l-1/2
         \end{array} \right.
\end{eqnarray}
The spin-orbital splitting is due to 
the corresponding  spin-orbital potential
\begin{equation}
   \displaystyle \frac 1 {E + 2M- V_V + V_S} \frac {\kappa} r
        \frac {d(2M - V_V + V_S)} {dr}
\label{spp}
\end{equation}
with some proper normalization factor.
It is seen that the spin-orbital splitting in RMF 
is energy dependent and they depends on the derivative of 
the potential $2M- V_V + V_S$ as well as the particle 
distribution. Therefore we introduce the so-called 
spin-orbital potential:
$V_{ls} = \displaystyle \frac {\kappa} r \frac {d(2M - V_V + V_S)} {dr}$. 
The $V_{ls}$ for $^{110}$Sn, $^{140}$Sn, and $^{170}$Sn,  
are given in Fig.3. The $V_{ls}$ for both proton and neutron 
are almost the same,
as the potential $V_V-V_S$ is a big quantity ( $\sim 700$ MeV ), 
the isospin dependence in the spin-orbital potential could 
be neglected. Therefore the proton and neutron $V_{ls}$ 
are the same in the present model. 
That is the reason why the spin-orbital splitting
for the neutron and proton is very close to each other in Fig. 2. 
From $^{110}$Sn to $^{170}$Sn, the amplitude of $V_{ls}$ 
decreases monotonically due to the surface diffuseness.

For the decline of the spin-orbital splitting, it come from 
the diffuseness of the potential or the outwards tendency of the 
potential. The diffuseness of the neutron potentials 
$V_V + V_S $ are given 
in Fig. 4 for $^{110}$Sn, $^{120}$Sn, $^{130}$Sn, $^{140}$Sn, $^{150}$Sn,
$^{160}$Sn, and $^{170}$Sn. It is seen that the depth of the 
potential decreases monotonically from the proton drip line to the
neutron drip line and the surface of the potential moves outwards.
The inserted figures gives the radii $R_0$ at which $V_V + V_S= -10$ MeV 
as a function of the mass number. 
The proton potentials $V_V + V_S$ 
for $^{110}$Sn,$^{120}$Sn, $^{130}$Sn,  $^{140}$Sn,$^{150}$Sn,
$^{160}$Sn,  and $^{170}$Sn are given in Fig. 5. With the increase
of the neutron number, the proton potentials are pushed towards outside 
as well due to the proton-neutron interaction. 

Similar neutron potentials $V_V - V_S$
for $^{100}$Sn, $^{110}$Sn,$^{120}$Sn, $^{130}$Sn,  $^{140}$Sn,$^{150}$Sn,
$^{160}$Sn,  and $^{170}$Sn are given
in Fig.6  and its inserted figures gives the radii $R_0$ at which 
$V_V - V_S=100$ MeV as a function of the mass number.
As seen in the above equations, the spin-orbital splitting is 
related with the derivative of the potential 
$V_V - V_S$ .
The surface diffuseness happens for both the vector and 
scalar potential, it occurs in  $V_V - V_S$ 
and  $V_V + V_S$.

So far we have started the isospin dependence and showed how the 
spin-orbital splitting is reduced due to the diffuseness of the 
potential. Another feature of the spin-orbital splitting is the 
energy-dependence given in Eq.(\ref{larspinor4}). 
The  single particle levels
in the canonical basis for the neutron in
$^{170}$Sn are given in Fig.7. 
The neutron potentials $V_V + V_S$
is represented by the solid line and the Fermi level 
is represented by a dashed-line. 
The depth of the neutron potential in 
$^{170}$Sn is about -60 MeV and the single particle 
levels begins from -55 MeV for $1s_{1/2}$. The spin-orbital 
partners appear in the order of the energy ($1p_{3/2}, 1p_{1/2}$), 
($1d_{5/2}, 1d_{3/2}$), ($1f_{7/2}, 1f_{5/2}$), 
($2p_{3/2}, 2p_{1/2}$), ($1g_{9/2}, 1g_{7/2}$),
($2d_{5/2}, 2d_{3/2}$), ($1h_{11/2}, 1h_{9/2}$), 
($2f_{7/2}, 2f_{5/2}$), ($3p_{3/2}, 3p_{1/2}$), and
(1$i_{13/2}, 1i_{11/2}$), etc. The larger spin-orbital
splitting for the same $l$ occurs when the particle 
is located in the middle of the potential. 
As it can be seen, the 
spin-orbital splitting in ($2p_{3/2}, 2p_{1/2}$) partners 
is twice as large as that in ($1p_{3/2}, 1p_{1/2}$) and thrice 
as large as that in ($3p_{3/2}, 3p_{1/2}$). 
Their energy dependence 
is clearly seen. In Fig. 8, the
spin-orbit splitting in Eq.(\ref{E-ls})
versus the binding energy in Eq.(\ref{abe}):
in $^{170}$Sn are also given in Fig.7 for
($p_{3/2}, p_{1/2}$), ($d_{5/2}, d_{3/2}$),
($f_{7/2}, f_{5/2}$), ($g_{9/2}, g_{7/2}$), 
($h_{9/2}, h_{11/2}$), and 
($i_{11/2}, i_{13/2}$) partners. 
It gives a strong energy dependence of the spin-orbital
splitting. Similar pattern is also appeared  
in other tin isotopes, but here we just choose $^{170}$Sn 
as an example.

As mentioned above, the
spin-orbital splitting is mainly due to
the corresponding  spin-orbital potential Eq.(\ref{spp}),
which decreases with the diffuseness of the potential. 
However, for the single particle levels in the same nuclei, 
the spin-orbital potential is the same, the difference is 
due to the particle distribution relative to the 
spin-orbital potential. In Fig. 9,  the spin-orbital potential
$V_{ls}$ multiplied by the density distributions for $p_{1/2}$ orbital 
in $^{170}$Sn 
is given in the upper panel, and 
the spin-orbital potential and the
probability distribution of the $p$-wave in $^{170}$Sn 
is given in the lower panel.
As seen in the lower panel, 
the particle distributions move outside with the 
increasing energy. The maximum overlap happens for 
($2p_{3/2}, 2p_{1/2}$). This clearly explain the features 
observed in Fig. 8. Another aspect in Fig.8 is that the splitting 
for $n=1$ always increases with $l$. This is because 
the centrifugal barrier keeps the particle away from center 
and prevent it from running away so that a big overlap 
always happens for larger $l$.

Recent developments in  accelerator technology and
detection techniques around the world
have made it possible to produce and study the
nuclei far from the stability line -- so called "EXOTIC
NUCLEI". Experiments of this kind have brought about
a new perspective to
nuclear structure physics: e.g., the neutron
halo and neutron skin as manifested in the rapid increase
in the measured interaction cross-sections in the
neutron-rich light nuclei \cite{TA.95,HJJ.95}. These extremely neutron-rich
nuclei and physics connected with the low density matter
in the tails of  the neutron and proton distributions have
attracted a lot of attention
in nuclear physics as well as in other fields such as astrophysics.
It is therefore of great importance to look into
the matter distribution and see
how the densities change with the  proton to neutron
ratio in these nuclei.
As the density here is obtained from a fully microscopic and
parameter free model, it is well supported by the experimental
binding energies.
We now proceed to examine the density distributions
of the whole chain of tin isotopes and study the relation between
the development of halo and shell effects within the model.
The density distributions for $^{100}$Sn, $^{110}$Sn,$^{120}$Sn, $^{130}$Sn,  
$^{140}$Sn,$^{150}$Sn, $^{160}$Sn,  and $^{170}$Sn
are given for neutron, proton, and matter density 
in Figs. 10a, 10b, and 10c, respectively. 
With the increase of $N$,
the density extends  toward outside, only a 
small increase is observed in the center. 
For the proton densities in Fig.10b, the density for $r \le4 $ fm 
decreases dramatically, 
and in reward for this a strong increase in the surface is observed 
with the increasing of the neutron number, because the 
proton number is fixed. This
is quite different from the case of Na isotopes, where
the proton distribution doesn't vary so much with $N$ 
in the surface area \cite{MTY.97}.
The matter densities for Sn isotopes are given in Fig.10c. 
Although neutron density increases in the center, it 
is slower than the decrease in proton. Therefore 
the total matter density decreases in the center. 
Due to the proton and neutron interaction, both 
proton and neutron densities increase in the surface. 

Fig. 11 is the same as Fig. 10, but in logarithm scale in order to 
see the detailed information in the tail.
With the increase of $N$,
the neutron density extends towards outside. 
The increase of the neutron density beyond $r = 6$ fm 
shows a gradual variation, except after $^{160}$Sn 
where the tail of the neutron density does not change.
The isotopes $^{160-170}$Sn have almost the same tail in the
density.  For the proton, the tail of the proton density in 
$^{100}$Sn and $^{110}$Sn shows difference from the 
isotopes with $N > 60$, which clearly shows that 
the tail is a asymptotic behavior mainly determined
by the last filled few particles. For the tail of the 
matter distribution in neutron-rich nuclei, 
they are mainly determined by the neutron density.

\section{Conclusions}

The Relativistic Continuum 
Hartree Bogoliubov ( RCHB ) theory, which is 
the extension of the
Relativistic Mean Field and the Bogoliubov transformation
in the coordinate representation, has been 
used to study the tin isotopes. 
The pairing correlation is
taken into account by a density-dependent force of zero
range. RCHB is used to describe the even-even tin 
isotopes all the way from the proton drip line
to the neutron drip line. 
The theoretical $S_{2n}$ as well as 
the neutron, proton, and matter $rms$ radii are presented 
and compared with the experimental values where they exist. 
The change of the potential surface with the 
neutron number has been investigated. 
The diffuseness of the potentials in tin isotopes is analyzed through the
spin-orbital splitting in order to provide new way
to understand the halo phenomena in exotic nuclei.
Summarizing the present investigation We can conclude: 
\begin{enumerate} 
\item The spin-orbital splitting decreases monotonically 
      from the proton drip line to the
      neutron drip line for all $l$ partners. 
      The spin-orbital splitting
      for the neutron and proton is very close to each other in RCHB,
      as the potential $V_V-V_S$ is a big quantity ( $\sim 700$ MeV ) and its 
      isospin dependence in the spin-orbital potential could
      be neglected. Therefore the proton and neutron $V_{ls}$
      are the same in the present model.
\item For the decreasing of the spin-orbital splitting with $N$, it come from
      the diffuseness of the potential or the outwards push of the
      potential. The depth of the
      potential decreases monotonically from the proton drip line to the
      neutron drip line and the surface of the potential moves outwards.
      With the neutron number $N$, the proton potentials 
      are pushed towards outside
      as well due to the proton-neutron interaction. 
\item A strong energy dependence of the spin-orbital
      splitting appears in tin isotopes.
      It is due to the change of wave function.
      The maximum overlap between the particle distributions 
      and the spin-orbital potential happens when 
      the single particle energy is in the middle of the potential. 
      This energy dependent behavior is stronger for 
      lower $l$ orbits than the higher $l$ orbits, 
      as the centrifugal barrier in higher $l$ orbits 
      keeps the particle away from center
      and prevent it from running away so that there is always 
      a reasonable overlap
      with spin-orbital potential.
\item With increase of $N$,
      the neutron density extends towards outside.
      The increasing of the neutron density beyond $r = 6$ fm
      shows a gradual variation with N.
      The neutron tail reaches saturation after $^{160}$Sn.
      For the proton, the tail of the proton density in
      $^{100}$Sn and $^{110}$Sn shows difference from the
      isotopes with $N > 60$, which clearly shows that
      the tail is a asymptotic behavior mainly due to 
      the last filled few particles. For the tail of the
      matter distribution in neutron-rich nuclei,
      they are mainly determined by the neutron density.
\item With the increase of $N$,
      the density extends  toward outside, only a
      small increase is observed in the center.
      For the proton, the density for $r \le 4 $ fm
      decreases dramatically,
      and in return for this a strong increase in the surface is observed
      with the increasing of the neutron number, because the
      proton number is fixed. Therefore, a slight decrease 
      in the center of the total matter density is observed. 
\end{enumerate}


\begin{table}
\caption{The neutron, proton, matter, and charge radii for Sn isotopes
\label{table}}
\begin{tabular}{cccccc||ccccc}
A   & $r_n$ & $r_p$ & $r_m$ & $r_c$ &&   A & $r_n$ & $r_p$ & $r_m$ & $r_c$\\ 
\tableline
 96 & 4.254 & 4.393 & 4.327 & 4.465 && 140 & 5.154 & 4.712 & 5.000 & 4.779 \\
 98 & 4.288 & 4.393 & 4.342 & 4.465 && 142 & 5.199 & 4.730 & 5.039 & 4.797 \\
100 & 4.320 & 4.393 & 4.357 & 4.465 && 144 & 5.243 & 4.748 & 5.077 & 4.815 \\
102 & 4.373 & 4.414 & 4.393 & 4.486 && 146 & 5.287 & 4.765 & 5.114 & 4.831 \\
104 & 4.422 & 4.433 & 4.428 & 4.505 && 148 & 5.331 & 4.781 & 5.152 & 4.847 \\
106 & 4.469 & 4.452 & 4.461 & 4.523 && 150 & 5.375 & 4.797 & 5.189 & 4.863 \\
108 & 4.515 & 4.470 & 4.494 & 4.541 && 152 & 5.419 & 4.812 & 5.227 & 4.878 \\
110 & 4.556 & 4.487 & 4.525 & 4.558 && 154 & 5.463 & 4.826 & 5.264 & 4.892 \\
112 & 4.601 & 4.502 & 4.557 & 4.573 && 156 & 5.506 & 4.839 & 5.301 & 4.905 \\
114 & 4.643 & 4.517 & 4.588 & 4.588 && 158 & 5.549 & 4.853 & 5.338 & 4.918 \\
116 & 4.684 & 4.532 & 4.619 & 4.602 && 160 & 5.590 & 4.866 & 5.374 & 4.931 \\
118 & 4.721 & 4.545 & 4.647 & 4.615 && 162 & 5.628 & 4.879 & 5.408 & 4.944 \\
120 & 4.760 & 4.559 & 4.677 & 4.629 && 164 & 5.661 & 4.894 & 5.439 & 4.959 \\
122 & 4.797 & 4.572 & 4.706 & 4.642 && 166 & 5.690 & 4.910 & 5.467 & 4.974 \\
124 & 4.833 & 4.586 & 4.735 & 4.655 && 168 & 5.715 & 4.926 & 5.492 & 4.991 \\
126 & 4.868 & 4.599 & 4.763 & 4.668 && 170 & 5.739 & 4.944 & 5.517 & 5.008 \\
128 & 4.902 & 4.611 & 4.790 & 4.680 && 172 & 5.761 & 4.961 & 5.541 & 5.025 \\
130 & 4.934 & 4.624 & 4.817 & 4.692 && 174 & 5.783 & 4.979 & 5.564 & 5.043 \\
132 & 4.964 & 4.636 & 4.842 & 4.704 && \\
134 & 5.013 & 4.656 & 4.883 & 4.724 && \\
136 & 5.061 & 4.675 & 4.923 & 4.743 && \\
138 & 5.108 & 4.694 & 4.962 & 4.761 && \\
\end{tabular}
\end{table}

\leftline{\Large {\bf Figure Captions}}
\parindent = 2 true cm
\parskip 1 cm
\begin{description}

\item[Fig. 1] Two-neutron separation energies $S_{2n}$ 
of even Sn isotopes as a function of $N$, 
including the experimental data ( solid points ) and the RCHB 
calculation with $\delta$-force ( open circles ).

\par

\item[Fig. 2] The neutron spin-orbit splitting
$E_{ls} = \displaystyle \frac {E_{lj=l-1/2}-E_{lj=l+1/2}} {2l+1}$
versus the mass number $A$ 
in tin isotopes for neutron ($1d_{3/2}, 1d_{5/2}$), 
($1g_{7/2}, 1g_{9/2}$), ($1i_{11/2}, 1i_{13/2}$), 
($1p_{1/2}, 1p_{3/2}$), ($1f_{5/2}, 1f_{7/2}$), 
and ($1h_{9/2}, 1h_{11/2}$) orbital and proton 
($1d_{3/2}, 1d_{5/2}$), and ($1f_{5/2}, 1f_{7/2}$) 
orbital, respectively.

\par

\item[Fig. 3] The derivative of the neutron potentials
$V_V( r ) - V_S( r )$
for $^{110}$Sn, $^{140}$Sn, and $^{170}$Sn.

\par

\item[Fig. 4] The neutron potentials $V_V( r ) + V_S( r )$ 
for $^{100}$Sn, $^{110}$Sn,$^{120}$Sn, $^{130}$Sn,  $^{140}$Sn,$^{150}$Sn,
$^{160}$Sn,  and $^{170}$Sn. In order to examine the surface diffuseness more 
clearly, the radii $R_0$ at which $V_V (R_0) + V_S(R_0) = -10$ MeV 
has been given as an inserted figure.
\par

\item[Fig. 5] The proton potentials $V_V( r ) + V_S( r )$ 
for $^{110}$Sn,$^{120}$Sn, $^{130}$Sn,  $^{140}$Sn,$^{150}$Sn,
$^{160}$Sn,  and $^{170}$Sn.  In order to examine the surface diffuseness more
clearly, the radii $R_0$ at which $V_V (R_0) + V_S(R_0) = -10$ MeV 
has been given as an inserted figure.

\par

\item[Fig. 6] The neutron potentials $V_V( r ) - V_S( r )$
for $^{100}$Sn, $^{110}$Sn,$^{120}$Sn, $^{130}$Sn,  $^{140}$Sn,$^{150}$Sn,
$^{160}$Sn,  and $^{170}$Sn. In order to examine the surface diffuseness more
clearly, the radii $R_0$ at which $V_V(R_0) - V_S(R_0) = 100$ MeV 
has been given as an inserted figure.

\par

\item[Fig. 7]  The  single particle levels
in the canonical basis for the neutron in
$^{170}$Sn. The neutron potentials $V_V( r ) + V_S( r )$
is represented by the solid line and the Fermi level 
is represented by a dashed-line. 
\par

\item[Fig. 8] The energy dependence of neutron spin-orbital
splitting in $^{170}$Sn,
\par

\item[Fig. 9] The upper panel is the spin-orbital potential
$V_{ls}$ multiplied by the density distributions of the $p_{1/2}$ orbital.
The lower panel is the spin-orbital potential and the
density distributions for $p_{1/2}$ orbital in $^{170}$Sn.
\par

\item[Fig. 10] The neutroni (a) , proton (b), and matter (c)
density distributions in Sn isotopes. 
\par

\item[Fig. 11] The same as Fig. 10, but in logarithm scale.

\end{description}


\begin{references}
\bibitem{THH.85b} I. Tanihata et al, Phys. Rev. Lett. {\bf 55} (1985) 2676

\bibitem{MR.96} J. Meng, and P. Ring, Phys. Rev. Lett. {\bf 77}
    (1996) 3963.

\bibitem{MR.97} J. Meng, and P. Ring, Phys. Rev. Lett. {\bf 80}
    (1998) 460.

\bibitem{ME.98} J. Meng, Nucl. Phys. {\bf A635} (1998) 3.

\bibitem{Zhu.93} M. Zhukov et al, Phys. Rep. {\bf 231} (1993) 151   

\bibitem{DFT.84}J.Dobaczewski, H.Flocard and J.Treiner, Nucl.Phys.
    {\bf A422} (1984) 103.

\bibitem{Ter.96} J. Terasaki, et al ,
    Nucl. Phys. {\bf A600} (1996) 371

\bibitem{KUO.97} T.T.S.Kuo, et al, Phys. Rev. Lett. {\bf 78}
    (1997) 2708

\bibitem{MTY.97} J. Meng, I. Tanihata and S. Yamaji, Phys. Lett. {\bf B419}
(1998) 1.

\bibitem{BB.77}J. Boguta, A.R. Bodmer, Nucl. Phys. {\bf A292} (1977) 413  

\bibitem{BGG.84}J.F. Berger et al, Nucl. Phys. {\bf A428} (1984) 32c  

\bibitem{SNR.93}M.M. Sharma, M.A. Nagarajan, and P. Ring,
    Phys. Lett. {\bf B312} (1993) 377

\bibitem{MSY.98} J. Meng, K.Sugawara-Tanabe, S.Yamaji, P. Ring and A.Arima,
Phys. Rev. C58 (1998) R628

\bibitem{TA.95} 
    I.Tanihata, Prog.Part.and Nucl.Phys.,{\bf 35} ( 1995 ) 505

\bibitem{HJJ.95} P.G. Hansen, A.S. Jensen, and B. Jonson,
    Ann. Rev. Nucl. Part. Sci. {\bf 45} (1995) 591

\end{references}
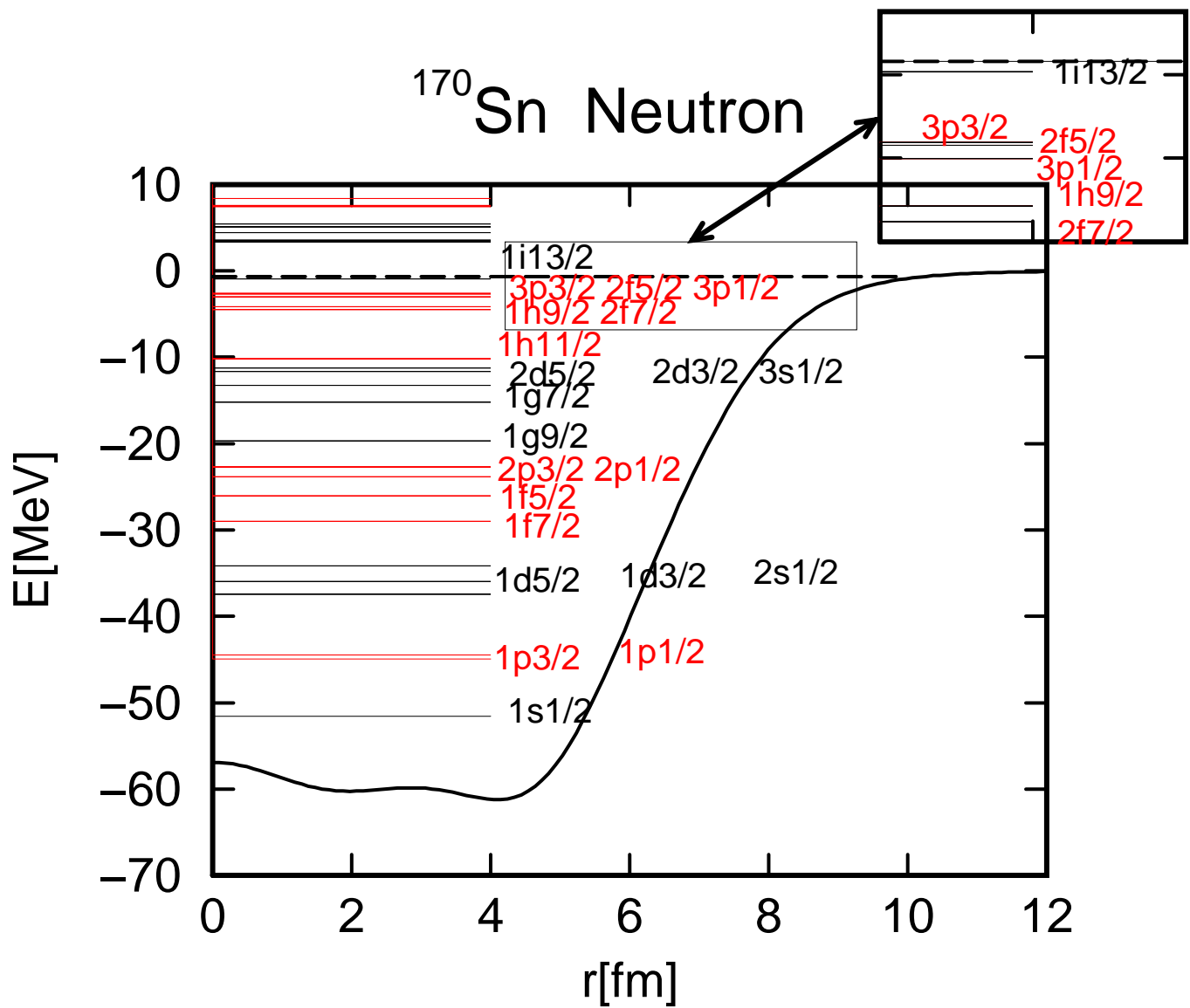
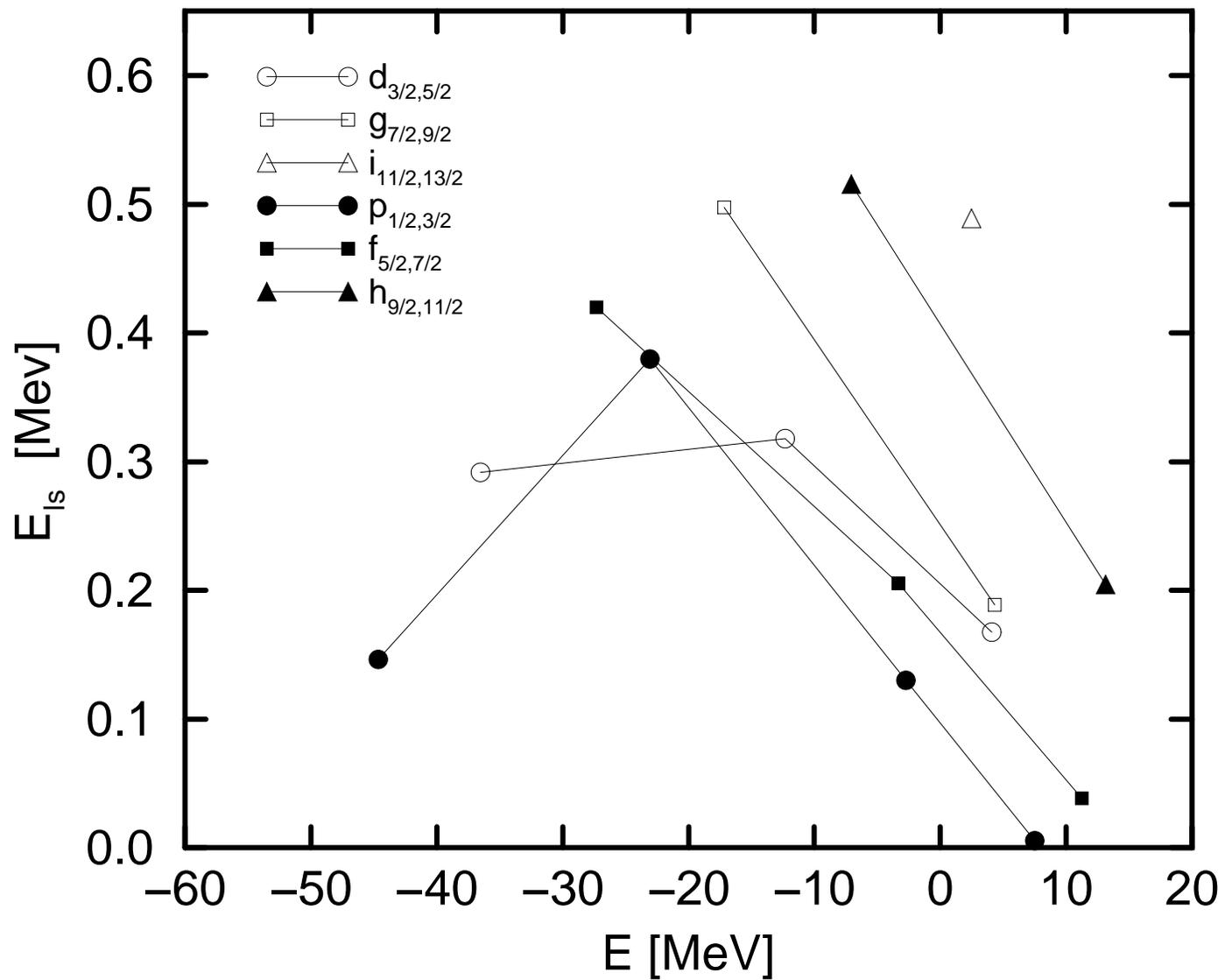
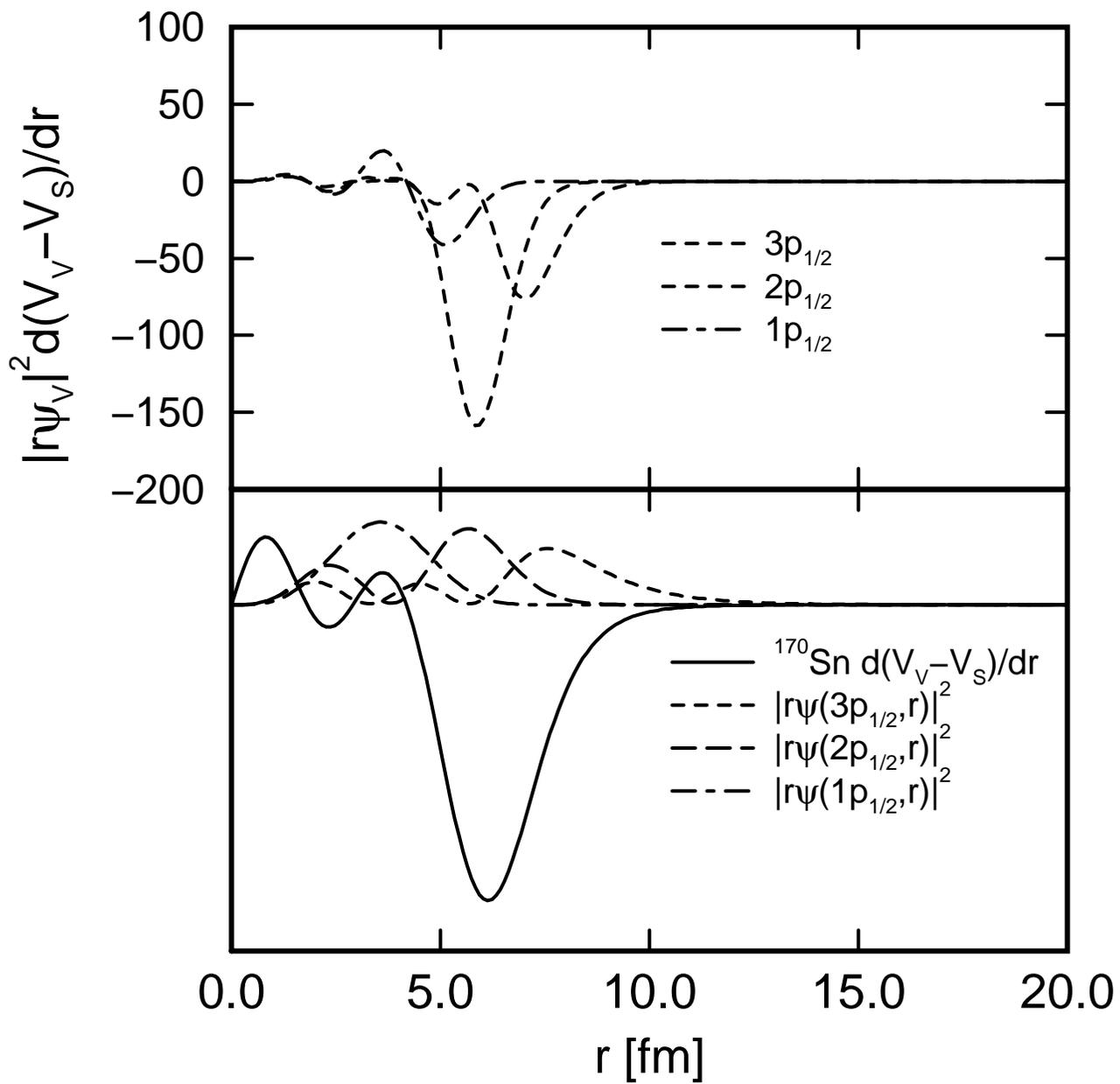
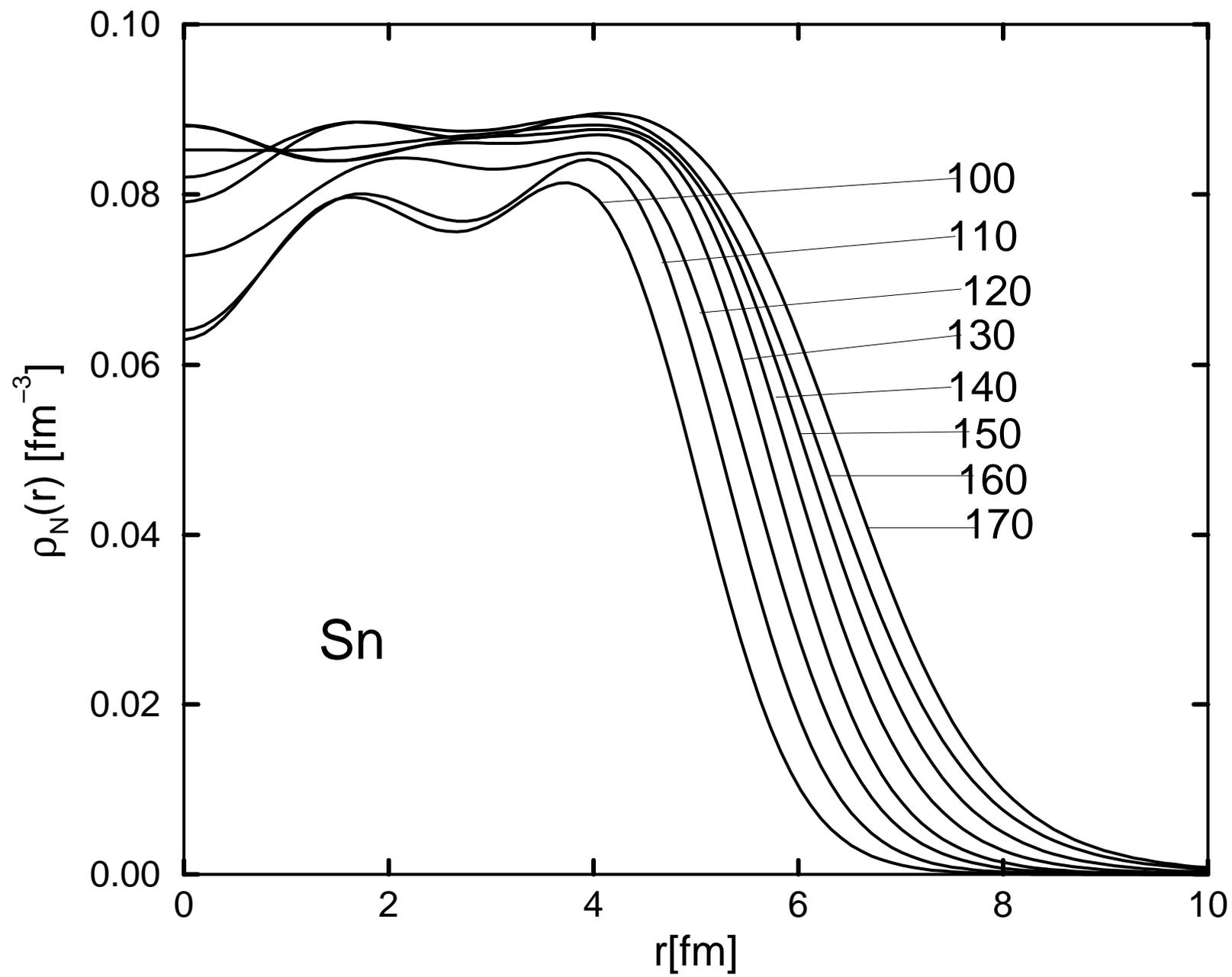
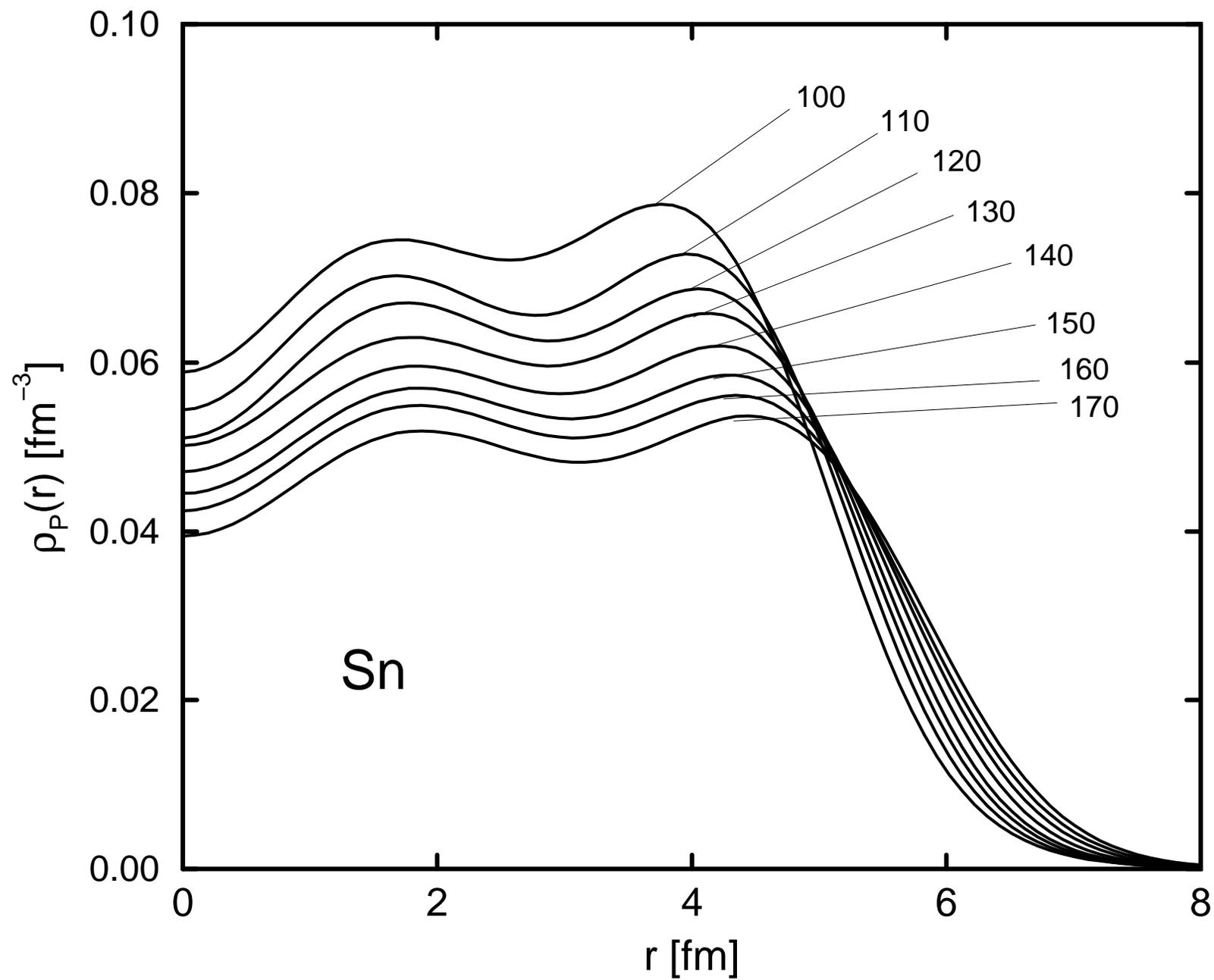
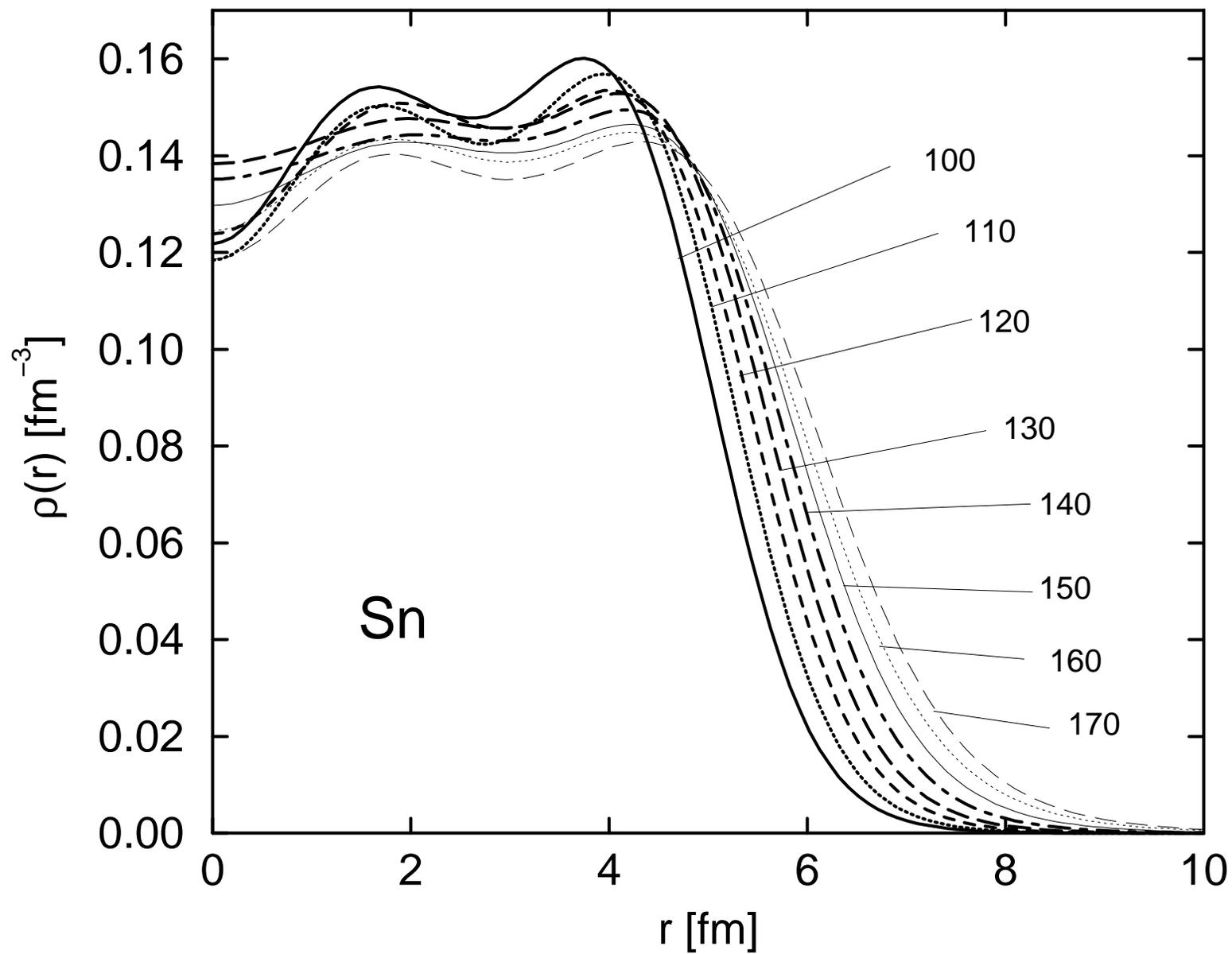
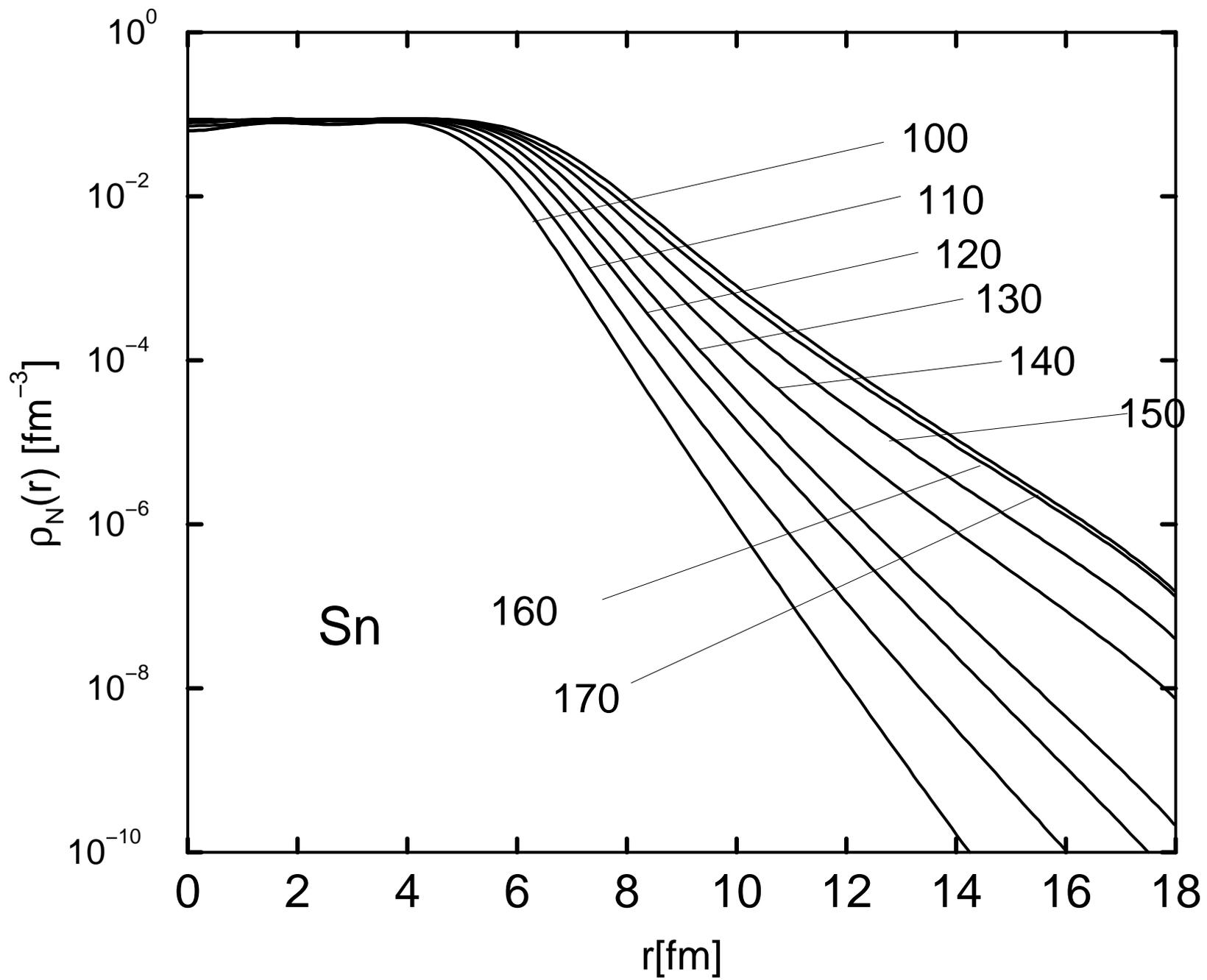
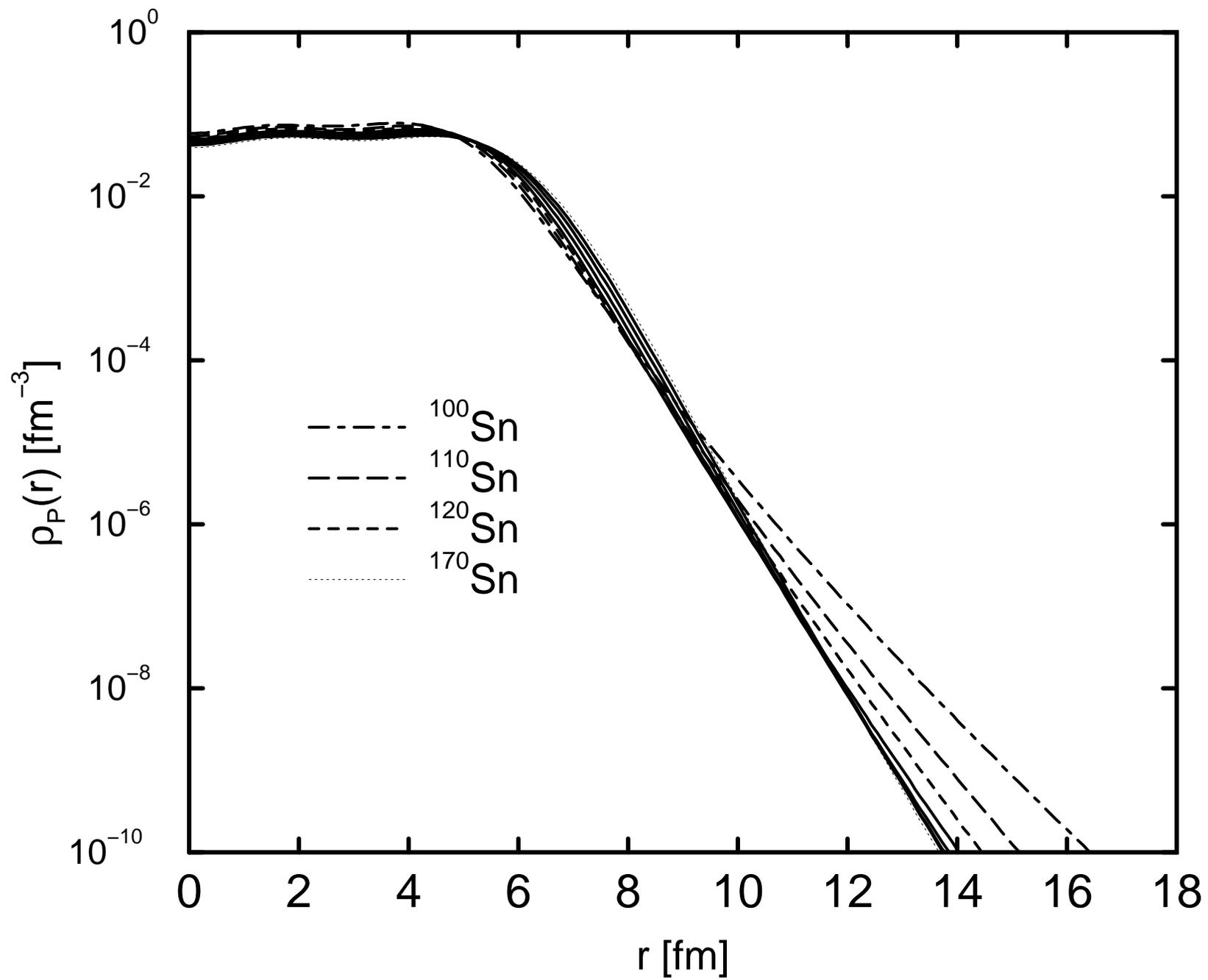
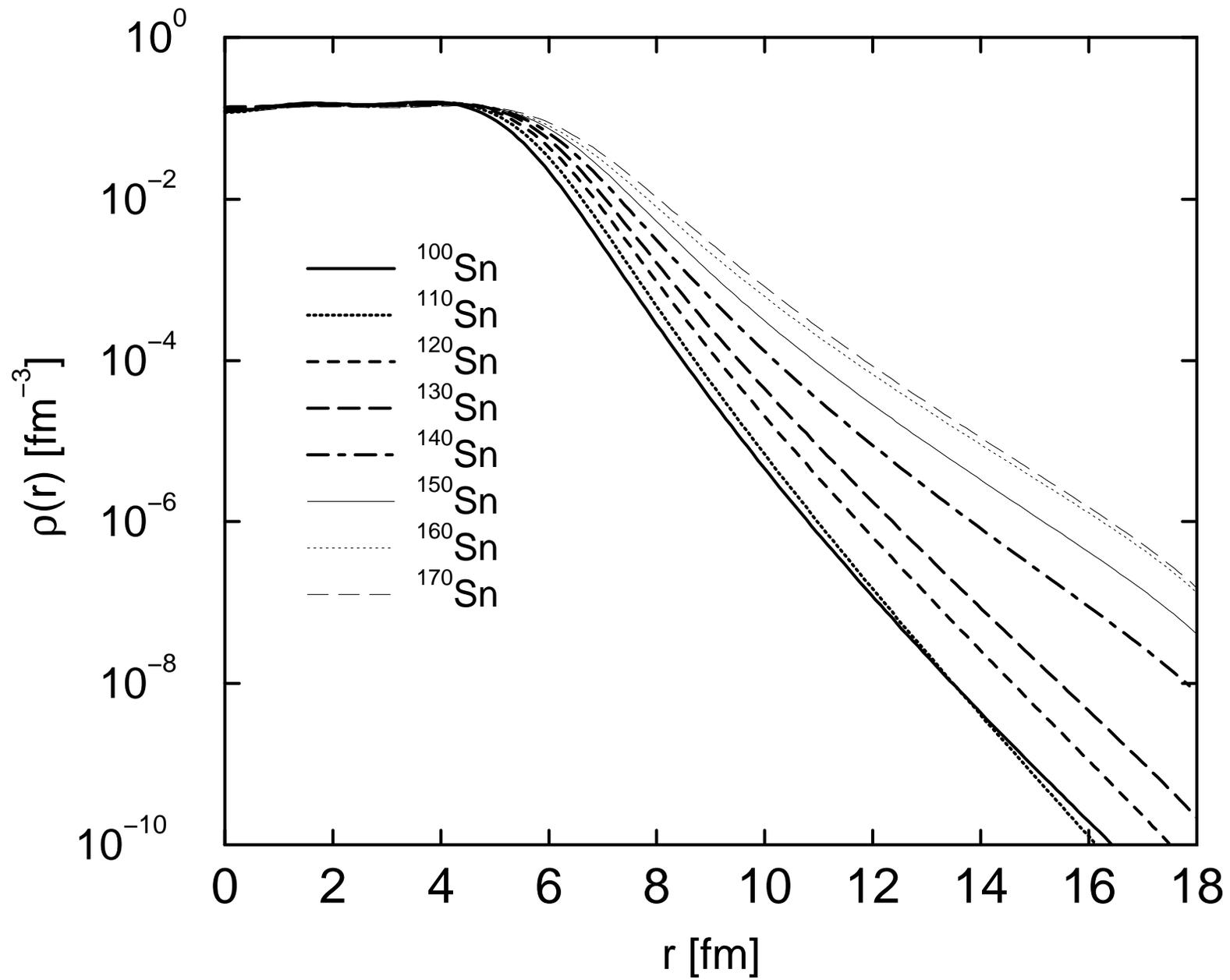
\end{document}